\documentclass[journal]{IEEEtran}

\usepackage{amssymb, amsmath}
\usepackage{amsthm}
\usepackage{graphicx}

\usepackage{algorithm}
\usepackage[noend]{algpseudocode}

\newcommand{\nop}[1]{}

\newtheorem{definition}{Definition}

\newtheorem{theorem}{Theorem}
\newtheorem{corollary}{Corollary}
\newtheorem{lemma}{Lemma}
\newtheorem{proposition}{Proposition}

\begin{document}

\title{Network Calculus Bounds for Time-Sensitive Networks: A Revisit} 

\author{Yuming Jiang \\NTNU, Norwegian University of Science and Technology, Trondheim, Norway}      
\maketitle

\begin{abstract}

Network calculus (NC), particularly its min-plus branch, has been extensively utilized to construct service models and compute delay bounds for time-sensitive networks (TSNs). This paper provides a revisit to the fundamental results. 
In particular, counterexamples to the most basic min-plus service models, which have been proposed for TSNs and used for computing delay bounds, indicate that the packetization effect has often been overlooked. 
To address, the max-plus branch of NC is also considered in this paper, whose models handle packetized traffic more explicitly. 
It is found that mapping the min-plus models to the max-plus models may bring in an immediate improvement over delay bounds derived from the min-plus analysis. In addition, an integrated analytical approach that combines models from both the min-plus and the max-plus NC branches is introduced. In this approach, the max-plus $g$-server model is extended and the extended model, called $g^{x}$-server, is used together with the min-plus arrival curve traffic model. By applying the integrated NC approach, service and delay bounds are derived for several settings that are fundamental in TSNs. 
\end{abstract}


\section{Introduction}\label{sec-1}

Time-Sensitive Networking (TSN) is a new IEEE standard that allows switches in local area networks (LANs) to support performance guarantees to time-sensitive applications, such as real-time audio/video (AV) data streams \cite{8021Q}\cite{8021Qcr}\cite{8021AS}\cite{8021BA}. Specifically, TSN uses priority and controlled queue draining algorithms for transmission selection at switch ports. To date, a number of transmission selection or queue draining schemes have been specified or recommended \cite{8021Q}\cite{8021Qcr}\cite{8021AS}\cite{8021BA}. Among them, Strict Priority (SP) is the default. In addition, Credit-Based Shaper (CBS) and its use with SP are specified in \cite{8021Q} and \cite{8021BA} to support AV streams. Other transmission selection schemes, which may be used together with SP and CBS, include Enhancements for Scheduled Traffic (EST) specified in \cite{8021Q}, (not specified) Enhanced Transmission Selection (ETS) \cite{8021Q}, and Asynchronous Traffic Shaping (ATS) \cite{8021Qcr}. A closely related and emerging standard for the Internet is IETF Deterministic Networking (DetNet) \cite{DetNet}. 

Network calculus is a queueing theory for performance guarantee analysis of communication networks \cite{Chang00}\cite{NetCal}\cite{SNC}\cite{DNC}. A key idea of network calculus is to model the traffic and service processes using some bounding functions and base the analysis on them. To this aim, the min-plus algebra and the max-plus algebra are exploited in the modeling and analysis. Accordingly, the network calculus theory has two branches --- the min-plus branch and the max-plus branch. 
In the TSN literature, the min-plus NC branch has been extensively utilized to construct service models and compute delay bounds. Representative results include \cite{MB14}, \cite{Zhao18}, \cite{Ehsan18} and \cite{Zhao21}. A comprehensive review can be found in \cite{Zhao22}, and a more recent effort of trying to improve network calculus nodal delay bounds for TSNs is \cite{Ehsan23}. All these results rely on the min-plus branch to model the service of the studied transmission selection schemes and compute delay bounds. In the recent work \cite{Ehsan23}, a traffic model from the max-plus branch is exploited to improve the delay bounds. 

In this paper, we first provide a revisit to the most fundamental NC results for TSNs, namely service curve and delay bounds offered by a link, a queue in a priority system, and a queue whose draining is controlled by CBS. This is motivated by that in TSNs, the latency from one device, e.g. Bridge A, to the next device, Bridge B, ``is measured from {\em arrival of the last bit} at (a port,) Port $n$ of Bridge A to the {\em arrival of the last bit} at (a port, ) Port $m$ of Bridge B'' \cite{8021Q}. However, counterexamples to the currently used service curve models for a link, an SP queue, and a CBS queue can be constructed, indicating that the packetization effect has been overlooked in those models. This consequently puts forward a question about the validity of the delay bounds computed from the service curve models that have overlooked the packetization effect. 

To address the concern, we examine the delay definitions in the two branches of network calculus. We prove that computing delay as the difference between the (last bit) arrival times is not only most direct and intuitive but also may lead to tighter delay bounds. Accordingly, two approaches are introduced for delay bound analysis. One is to use the max-plus network calculus models as the basis and map the min-plus models to them. The other is an integrated approach, where for traffic, the min-plus arrival curve model is used, while for service, an extension, called $g^{x}$-server, to the max-plus $g$-server model is proposed. While the first approach does lead to improved delay bounds, the improvement is not enough to match with the min-plus based delay bounds that ignore the packetization effect in the service curve models for the counterexample cases. On the other hand, the second approach successfully recovers or even improves the delay bounds. Finally, the integrated approach is applied to several typical settings with SP and CBS, for which, bounds for their service and delay are proved. 
The key contributions of this paper can be summarized as: 
\begin{itemize}
	\item A closer investigation on the packetization effect and its impact on some fundamental min-plus service models used in TSNs, cf. Propositions \ref{pr-pkt}, \ref{pr-sp} and \ref{pr-cbs};
	\item A revisit to the delay definition difference in the two NC branches and its implication on delay bound analysis, cf. Proposition \ref{pr-4}; 
	\item An approach for improving delay bounds by mapping the min-plus models to the max-plus models and computing delay bounds in the max-plus domain, cf. Lemma \ref{lm-map} and Theorem \ref{th-1}; 
	\item A integrated approach for delay bound analysis, based on the proposed max-plus $g^{x}$-server model, cf. Propositions \ref{pr-5} and \ref{pr-6} and Theorem \ref{th-2};
	\item Service and delay bounds for SP and CBS when used separately or in combination under different settings that are fundamental in TSNs, cf. Theorems \ref{th-sp} -- \ref{th-cbs4}.
\end{itemize}

The rest is organized as follows. In Section \ref{sec-2}, TSN transmission selection is introduced, together with the system model and notation. In Section \ref{sec-3}, the most basic traffic and server models and delay bounds from both branches of network calculus are introduced. In Section \ref{sec-4}, the packetization effect is discussed in combination of counterexamples to the fundamental service curve models that have been widely adopted in TSNs. In Section \ref{sec-5}, the two approaches for delay bound analysis are discussed. In Section \ref{sec-6}, the integrated analytical approach is applied to study SP and CBS under several settings and derive service and delay bounds for them. Finally concluding remarks are given in Sec.~\ref{sec-7}.

\section{Transmission Selection in TSNs, and System Model for Analysis} \label{sec-2} 

\subsection{TSN Transmission Selection Algorithms}

In TSNs, the transmission of frames, which will also be called packets in this paper, on a port of a switch is managed by transmission selection algorithms. Frames are transmitted on the basis of the traffic classes and the transmission selection algorithms supported by the corresponding queues \cite{8021Q}\cite{8021Qcr}\cite{8021BA}. 
To date, the TSN standard has specified the operations of three transmission selection algorithms, which are Strict Priority (SP), Credit-Based Shaper (CBS),  and Asynchronous Traffic Shaping (ATS) \cite{8021Q}\cite{8021Qcr}\cite{8021BA}. In addition, enhancements for scheduled traffic (EST) via timed transmission gate control are also specified \cite{8021Q}. These transmission selection schemes may be implemented and work together \cite{8021Q}\cite{8021Qcr}\cite{8021BA}. 

\subsubsection{Strict Priority (SP)}
In time-sensitive networking, (non-preemptive) SP is the default algorithm for transmission selection among queues of different traffic classes \cite{8021Q}. When a higher priority queue has packets, they will be selected for transmission before lower priority queues. When a packet with higher priority arrives seeing a lower priority packet under transmission, the transmission will not be preempted. 

\subsubsection{Credit Based Shaper (CBS)}
For controlled queue draining, credit-based shaping is specified in TSN \cite{8021Q}. A credit based shaper (CBS) has a parameter, called $idleShope$, which determines the fraction of the transmit data rate (in bps) of the port or link, called $portTransmitRate$, available to the CBS queue. Another parameter, called $sendSlope$, is also used in introducing the operation of CBS, which is set by default as $sendSlope = idleShope - portTransmitRate$. 
In addition, a counter, called $credit$, initialized to zero, is used during the operation. CBS operates as follows: 
\begin{itemize}
\item When the CBS queue is not empty and the value of $credit$ is non-negative, i.e. $credit \ge 0$, the head-of-queue packet, if the CBS queue is chosen by the scheduling algorithm, is transmitted. 
\item The value of $credit$ is decreased with the send slope $sendSlope$ {\em during the transmission of a packet from the CBS queue}. 
\item When there are packet(s) in the CBS queue waiting but none is being transmitted, e.g. due to the queue not selected for transmission by the scheduling algorithm or the value of $credit$ is negative, $credit$ is increased with the idle slope $idleSlope$. 
\item When the CBS queue becomes/is empty, if the value of $credit$ is positive, it is set to zero; otherwise, it is increased with $idleSlope$ until zero.  
\end{itemize}

\subsubsection{Asynchronous Traffic Shaping (ATS)}
In TSN, asynchronous traffic shaping, with the concept originally proposed in \cite{ECRTS16}, is used to support flows requiring bounded end-to-end latency but without the need of synchronizing transmissions at switches across the network \cite{8021Qcr}. Specifically, an asynchronous traffic shaper enforces the traffic of each flow sharing the queue, where CBS may also be implemented, to conform to its initial traffic specification at the entrance \cite{8021Qcr}. 
An appealing property of ATS is that appending an ATS shaper to a FIFO system will not increase the delay bound \cite{ECRTS16, LeBoudec18, Jiang22}. 

\subsubsection{Enhancements for Scheduled Traffic (ETS)}

ETS \cite{8021Q} specifies time-aware queue-draining procedures and extensions to enable TSN switches and end stations to schedule the transmission of frames based on timing derived from the standard \cite{8021AS}. 
Specifically, a transmission gate is associated with each queue and EST schedules transmission via timed gate control \cite{8021Q}: the state, open or closed, of the transmission gate determines whether or not frames from the queue can be selected for transmission in accordance with the transmission selection algorithms associated with the queue. {\em When used with CBS,  the $credit$ is accumulated only when the gate is open and put on hold when the gate is closed} \cite{8021BA}. 

\vspace{6pt}
\noindent 
Because of the ATS-shaping-for-free property \cite{ECRTS16, LeBoudec18, Jiang22}, a bound on the queueing related delay of the end-to-end latency can be easily obtained by adding bounds on the nodal delays when ATS is accordingly applied \cite{ECRTS16, LeBoudec18, Ehsan18}. For this reason, we will focus on nodal bounds for SP and CBS in this paper. Specifically, service and delay bounds for them under standalone and combined-use settings will be derived, where the effect of $credit$-holding on CBS due to ETS will be taken into consideration. 

\subsection{System Model}

We consider FIFO systems serving flows in a packet-switched time-sensitive network. Such a system may be a queue served by a link, a strict priority scheduler, a credit-based shaper, or a combination of them. By convention, {\em a packet is said to have arrived to (respectively served by) the system when and only when its last bit has arrived to (respectively departed from) the system} \cite{ 8021Q}. When a packet arrives, the packet may be queued and the buffer size for the queue is assumed to be large enough ensuring no packet loss. The queue is FIFO and initially empty. 

A flow is a sequence of packets that may arrive at a point in the system at different time instances. The flow starts from packet $n=1$, i.e. the 1-st packet. For convenience and compatibility with the notation used in \cite{Chang00}, packet 0 is defined to be a virtual packet with arrival time at $0$ and packet length $0$. In addition, for a negative packet number, a similar virtual packet is defined, which also has zero length but whose arrival time is negative. 

We use $A(t)$ to denote the cumulative traffic amount of the flow entering the system and $A^{*}(t)$ the cumulative amount of traffic output from the system, up to time $t$ (excluded). By convention, we adopt $A(0) = A^{*}(0) = 0$. In addition, we define $A(s,t) \equiv A(t)-A(s)$ and $A^{*}(s,t) \equiv A^{*}(t)-A^{*}(s)$, which respectively denote the amount of input traffic and the amount of output traffic in period $[s, t)$. Since traffic at $t$ is excluded, $A(t,t)=0$ by convention and so is $A^{*}(t,t)=0$.

Also we model the traffic processes of the flow by marked point processes.  Specifically, the input to the system is by a marked point process $(\overrightarrow{a}, \overrightarrow{l})$ which consists of two sequences of variables $\overrightarrow{a}=\{a(n), n=0, 1, 2, ...\}$ and $\overrightarrow{l}=\{l(n), n=0, 1, 2, ...\}$, where $a(n)$ denotes the arrival time of the $n$-th packet and $l(n)$ its length (in bits). For the output, a similar marked point process is defined which is $(\overrightarrow{d}, \overrightarrow{l})$  with $\overrightarrow{d} =\{d(n), n=0, 1, 2, ...\}$, where $d(n)$ denotes the departure time of the $n$-th packet from the system with $d(0)$ set to $0$. 

For the flow, define 
\begin{equation}
L(n) \equiv \sum_{m=0}^{n-1} l(m)
\end{equation}
and $L(m,n) \equiv L(n) - L(m)$.  Since $l(n)$ is not included in $L(n)$, by convention, $L(n,n) =0$ and $L(0) = 0$ . 

In addition, we use $l^{M}$ and $l^{m}$ to respectively denote the maximum packet length and the minimum packet length of the flow. When there are multiple priority queues, a subscript is added to differentiate. Specifically, $l^{M_l}$ and $l^{m_l}$ (resp. $l^{M_u}$ and $l^{m_u}$) denote the maximum and minimum packet length of lower priority queues (resp. higher priority queues). 

The following equation establishes the relation between the two ways of representing the traffic process, which can be verified from their definitions,
\begin{equation}
A(t) = \sum_{0\le m }l(m) \mathcal{I}_{a(m) < t}
\end{equation}
where the indicator function is defined as $\mathcal{I}_{a(m) < t} =1$ if $a(m) <t$, and 0,  otherwise. Similarly, $A^{*}(t) = \sum_{0\le m }l (m) \mathcal{I}_{d(m) < t}$. 

In this study, a focus is on finding delay bounds. The delay of a packet $n (\ge 1)$, denoted by $D(n)$, is:
\begin{equation}
D(n) = d(n) -a(n).
\end{equation}
In addition, the virtual delay at time $t (>0)$ is defined as 
\begin{equation}
D(t) = \inf\{\tau \ge0: A(t) \le A^{*}(t+\tau)\}. 
\end{equation}

\subsection{Additional Notation}

The set of nonnegative nondecreasing functions, denoted by  $\mathbf{\mathcal{F}}$, is defined as: 
$$\mathbf{\mathcal{F}} = \{f: 0 \le f(s) \le f(t), \forall s\le t \}$$ and  
$\mathbf{\mathcal{F}}_0$ its subset with $f(0) = 0$. By their definitions, $A(\cdot)$, $A^{*}(\cdot)$ and $L(\cdot)$ are all in $\mathbf{\mathcal{F}}_0$. 

For $f \in \mathbf{\mathcal{F}}$, its lower and upper pseudo-inverse functions, denoted as $f^{\downarrow}$ and $f^{\uparrow}$, are respectively defined as:
\begin{eqnarray}
f^{\downarrow}(y) &\equiv& \inf \{x\ge 0: f(x) \ge y\} \nonumber\\
f^{\uparrow}(y) &\equiv& \sup \{x\ge 0: f(x) \le y\} \nonumber
\end{eqnarray}
The pseudo-inverse functions have a number of properties \cite{Chang00} \cite{JL17}. One of them is that both $f^{\downarrow}$ and $f^{\uparrow}$ are nonnegative and nondecreasing, i.e., $f^{\downarrow}, f^{\uparrow} \in \mathbf{\mathcal{F}}$. 

For a variable $x$, we define:
\begin{eqnarray}
(x)^{+} &=& \max\{x, 0\} \nonumber\\
x^{+} &\equiv& x + \epsilon, \epsilon \to 0  \nonumber \\
x^{-} &\equiv& x - \epsilon, \epsilon \to 0  \nonumber 
\end{eqnarray}

The horizontal distance and vertical distance between two functions $f, g \in \mathbf{\mathcal{F}}$, denoted as $H(f, g)$ and $V(f, g)$, are respectively defined as: 
\begin{eqnarray}
H(f,g) &\equiv&  \sup_{x\ge 0} \inf\{y \ge 0: g(x+y)- f(x) \ge 0 \} \nonumber\\ 
V(f,g) &\equiv& \sup_{x\ge 0}\{g(x) - f(x)\} \nonumber
\end{eqnarray}

\nop{
 A summary of the notation is provided in Table \ref{tb1}. 

\begin{table}[htb]
\caption{Notation}
\label{tb1}
\begin{tabular}{|r|l|}
\hline
$n$& The $n$-th packet, $n=0, 1, 2, \dots$ \\ \hline
$a(n)$& Arrival time of packet $n$ \\ \hline
$A(t)$& Cumulative amount of arrival traffic up to time $t$ (excluded) \\ \hline
$A(s, t)$& $A(s,t) = A(t)-A(s)$ \\ \hline
$d(n)$& Departure time of packet $n$ \\ \hline
$A^{*}(t)$& Cumulative amount of departure traffic up to time $t$ (excluded) \\ \hline
$D(n)$& Delay of packet $n$, i.e. $D(n) = d(n)-a(n)$ \\ \hline
$D(t)$& Virtual delay at time $t$: $D(t) = \inf\{s\ge0: A(t) \le A^{*}(t+s)\}$ \\ \hline
\end{tabular}%
\end{table}
}

\section{Network Calculus Basics} \label{sec-3}

The network calculus theory has two branches --- the min-plus branch and the max-plus branch. 
While the former establishes models based on $A(t)$ and $A^{*}(t)$, the latter on $(\overrightarrow{a}, \overrightarrow{l})$ and $(\overrightarrow{d}, \overrightarrow{l})$. For the deterministic version of NC, focused in this paper, similar models have been introduced under different names and settings \cite{Chang00}\cite{NetCal}\cite{DNC}\cite{JL17}. In this paper, the min-plus part will follow the terminology and settings used in \cite{NetCal}, while the max-plus part follows \cite{Chang00}. 

\subsection{Min-Plus Network Calculus: Models and Delay Bound}\label{sec-nc-min}

\begin{definition}\label{def-ac}
A flow is said to have an arrival curve $\alpha \in \mathbf{\mathcal{F}}$, if for all  $0\le s \le t$ \cite{NetCal}, 
\begin{equation}
A(s, t) \le \alpha(t-s).
\end{equation}
or equivalently, for all $t \ge 0$, $A(t) \le A \otimes \alpha(t) \equiv \{A(s) + \alpha(t-s)\}$.
\end{definition} 

An example arrival curve type is the token-bucket arrival curve. Specifically, if a flow is constrained by a token bucket with parameters $(\sigma, \rho)$, where $\sigma (\ge l^{M})$ is the bucket size and $\rho$ the token generation rate, the flow has an arrival curve $\alpha(t) = \rho t +\sigma$. Note that, by definition, $A(t, t)=0$, so we can always set $\alpha(0)=0$ making  $\alpha$ in $\mathbf{\mathcal{F}_0}$ without violating the arrival curve definition.  
To ease expression in the remaining, this will be implicitly set if not specified, and we shall simply call $\alpha(t) = \rho t +\sigma$ the arrival curve. 

\begin{definition}\label{def-sc}
A system is said to provide a service curve $\beta \in \mathbf{\mathcal{F}_0}$, if for any time $t\ge 0$, there exists some time $s \in [0, t]$ such that \cite{NetCal}
\begin{equation}
A^{*}(t) \ge A(s) + \beta(t-s) 
\end{equation}
or equivalently, there holds for all $t \ge 0$, $A^{*}(t) \ge A \otimes \beta(t)$.
\end{definition}

An example service curve type is the latency-rate service curve. Specifically, a latency rate service curve with rate $R$ and latency term $T$ is $\beta(t)=R(t-T)^{+}$. 

\vspace{6pt}
\noindent {\bf Delay bound: }{\em If the input to a system has an arrival curve $\alpha$ and the system provides to the input a service curve $\beta$, the virtual delay at any time $t (\ge 0)$ is upper-bounded \cite{NetCal}: }
\begin{equation}\label{ab-db1} 
D(t) \le H(\alpha, \beta) \equiv D^{(\alpha, \beta)}.
\end{equation}

As an example, if the input has a token-bucket arrival curve $\alpha(t) = \rho t +\sigma$ and the system provides a latency-rate service curve $\beta(t)=R(t-T)^{+}$, with $\rho \le R$, the upper-bound on the virtual delay becomes $\frac{\sigma}{R}+T$. 

\subsection{Max-Plus Network Calculus: Models and Delay Bound}\label{sec-nc-max}

\begin{definition}\label{def-g1}
A flow is said to be $g$-regular, with $g \in \mathbf{\mathcal{F}}_0$, if for all $(0 \le) m \le n$, there holds  \cite{Chang00}
\begin{equation}
a(n) -a(m) \ge g(L(m, n)). 
\end{equation}
or equivalently $a(n) \ge a \bar\otimes_L g(n) \equiv  \max_{0 \le m \le n} \{a(m) + g(L(m, n)\}$. 
\end{definition}

Length Rate Quotient (LRQ), proposed in \cite{ECRTS16}, is the first algorithm for ATS (asynchronous traffic shaping) \cite{8021Qcr}. Consider a flow regulated by a LRQ shaper, where the regulation rate to the flow is $r$. The LRQ shaper ensures that the time gap between two packets $n$ and $n+1$ of the flow is not smaller than $\frac{l(n)}{r}$, $\forall n \ge 0$. From this, it is easily verified that, for all $0\le m \le n$, $a(n)  \ge a(m) + \sum_{k=m}^{n-1} \frac{l(k)}{r}$, so the flow is $g$-regular with $g(v) = \frac{v}{r}$. 

\begin{definition}\label{def-g2}
A system is said to be a $g$-server ($g \in \mathbf{\mathcal{F}}$) to the input $a(n)$, $n=1, \dots$, if for the output, there holds for all $n \ge 1$  \cite{Chang00}:
\begin{equation}\label{max-sc}
d(n) \le \max_{0 \le m \le n} \{a(m) + g(L(m, n))\} . 
\end{equation}
which can also be written as $d(n) \le a \bar\otimes_{L} g(n)$.
\end{definition}

As an example, it can be verified that a single-flow LRQ shaper with regulation rate $r$ to the input is a $g$-server with $g(v)=\frac{v}{r}$. The proof follows from the proof of Lemma 4 in \cite{Jiang22} and the start-time server model that is a special case of the $g$-server model \cite{Jiang03}. 

\vspace{6pt}
\noindent {\bf Delay bound: }{\em If the input is $g_1$-regular and the system is a $g_2$-server to the process, then for any packet $n \ge 1$, its delay $D(n)$ is upper-bounded \cite{Chang00}:} 
\begin{equation}\label{gg-db2} 
D(n) \le \sup_{v \ge 0}\{g_2(v) - g_1(v)\} = H(g_1, g_2) \equiv D^{(g, g)}. 
\end{equation}

For the LRQ examples above, it can be proved that, if the input is LRQ-regulated with rate $r_1$,  there is no delay when it passes through any single-flow LRQ shaper with rate $r_2 (\ge r_1)$ \cite{Jiang22}.

\section{Impact of Packetization on the Analysis}\label{sec-4}

\subsection{Packetization on a Link}
\begin{figure}[th!]
\centering
  \includegraphics[width=0.5\linewidth]{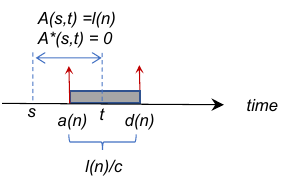} 
  \caption{Impact of packetization on the traffic and service of a link} 
  \label{fig-pkt}
\end{figure}

In TSNs, a packet is considered to have arrived (respectively been transmitted) when and only when its last bit has arrived (respectively been transmitted) \cite{ 8021Q}. To show its impact on the analysis, Figure \ref{fig-pkt} presents a simple example. 

Figure \ref{fig-pkt} illustrates the transmission of a packet $n$ on a link with rate $c$ and the amount of traffic or service in a time period $[s, t)$. If fluid traffic models were used, we would have $A(s,t)=A^*(s,t)=c(t-a(n))$. However, under the packet model, taking packetization into account, we actually have $A(s,t)=l(n)$ while $A^*(s,t)=0$. The former is because at time $a(n)$ that is within $[s,t)$, the (last bit of the) packet has arrived so that the transmission can start. On the contrary, for $A^*(s,t)$,  $d(n)$ is not within the period meaning the last bit of the packet has not finished transmission so the packet is not considered to have been transmitted or served in the considered period $[s,t)$. Because $0 \le c(t-a(n)) \le l(n)$, Proposition \ref{pr-pkt} is concluded. 

\begin{proposition}\label{pr-pkt}
For traffic on a link that has rate $c$, $c  t$ is neither an arrival curve of the traffic nor a service curve of the link. 
\end{proposition}

In the literature, it has been proved that the traffic on the link has an arrival curve $\alpha(t) = c t +l^{M}$ (see e.g. \cite{Jiang02}). In addition, since the arrivals are regulated by the link rate, they are LRQ-shaped with rate $c$. Hence, the traffic is $g_1$-regular with $g_1(v)=\frac{v}{c}$. Moreover, it has been proved that the link provides a service curve $\beta(t)=c \cdot (t- \frac{l^{M}}{c})^+$ \cite{Jiang03} and is a $g_2$-server with $g_2(v)=\frac{v+l^{M}}{c}$ to the traffic \cite{Chang00}\cite{Jiang03}.  

\subsection{Service Curve Models for TSNs: A Revisit}

Among the various TSN transmission selection algorithms, we focus on SP and CBS, which are the most fundamental ones. The former is the default algorithm for transmission selection among queues and CBS is specified for controlled queue draining on the queue it is applied. 
To examine their service models, we shall consider SP and CBS under the simplest settings. Specifically, for SP, we only consider the queue at the highest priority level, and for CBS, it is assumed to operate on a queue to which the link capacity is solely dedicated to. They are special cases of the studied settings in the TSN literature. For instance, the simplest SP setting is recovered by removing all higher priority traffic and the simplest CBS setting is retrieved by removing traffic from all other queues in related cases reviewed and/or studied in \cite{Zhao22} and references therein. 

Specifically, for the SP case, $ct$ has been used as the service curve for the link and sometimes also the service curve for the highest priority queue, e.g. in \cite{Zhao22} (cf. Eq. (9)) and references therein. Since, as summarized in Proposition \ref{pr-pkt}, considering packetization, $ct$ is not a service curve for the link, it cannot be a service curve for the highest priority queue sharing the link. Note that due to non-preemption, upon arrival, a highest priority packet will have to wait for the packet under transmission to complete even though it is from a lower priority queue. Factoring the fact, e.g. by adding $\frac{l^{M_l}}{c}$ at $a(n)$ in Figure \ref{pr-pkt}, similar analysis can be conducted, and we can conclude Proposition \ref{pr-sp}. 

\begin{proposition}\label{pr-sp}
For the highest priority queue sharing a link, which has rate $c$, with other queues using SP, neither $ct$ nor $c (t-\frac{l^{M_l}}{c})^{+} \equiv \beta^{SP*}(t)$ is a service curve. 
\end{proposition}

For the CBS queue at the highest priority level, a service curve widely used in the literature, e.g. in \cite{MB14, Zhao18, Ehsan18, Zhao21} and \cite{Zhao22} and references therein, is: 
$$ idleSlope (t-\frac{l^{M_l}}{c})^{+}.$$ 
For the simplest CBS case, where there is no lower priority queue, $l^{M_l}=0$, and hence it becomes $$ idleSlope \cdot t \equiv \beta^{CBS*}(t).$$ If this were true, we would have from the service curve definition: for all $t \ge 0$, 
$
A^{*}(t) \ge A\otimes \beta^{CBS*}(t) \equiv  \inf_{0 \le s \le t}\{A(s) + idleSlope \cdot (t-s)\}. 
$
In the following, a counter example is introduced. 

\begin{figure}[th!]
\centering
  \includegraphics[width=0.7\linewidth]{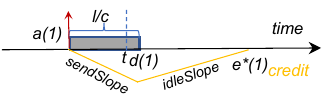} 
  \caption{Impact of packetization on CBS} 
  \label{fig-cbs-x}
\end{figure}

Consider a time instant $t \in (a(1), d(1))$ as illustrated in Figure \ref{fig-cbs-x}. The figure also shows the arrival time, departure time, the corresponding progression of $credit$ due to the transmission of the first packet that has length $l(1)=l$. 
It is clear from the figure that by $t$, only packet 1 has arrived and no packet has departed. Hence, considering packetization, $A(t)=l$ and $A^{*}(t)=0$. 
Now consider
$$
A\otimes \beta^{CBS*}(t) = \inf_{0 \le s \le t}\{A(s) + idleSlope (t-s)\}. 
$$
The right hand side can be divided into two parts. (1) $a(1) < s \le t$: In this case $A(s) = l$ and hence $A(s) + idleSlope (t-s) > l$. (2) $0 < s \le a(1)$. In this part, $A(s) = 0$ and the infimum is $idleSlope (t-a(1)) >0$. Together it can be concluded that, for the chosen $t$, there holds
$$
A\otimes \beta^{CBS*}(t) > 0 = A^{*}(t)
$$
which contradicts the service curve definition $A^{*}(t) \ge A\otimes \beta^{CBS*}(t)$. In summary, we have proved Proposition \ref{pr-cbs}. 

\begin{proposition}\label{pr-cbs}
For the single CBS queue on a link with rate $c$, $idleSlope \cdot t $ is not a service curve. 
\end{proposition}

\subsection{Implication}
The above investigation indicates that packetization may impact the NC-based TSN analysis significantly. As a remark, the effect of packetization has already been considered in the general network calculus framework. In particular, the min-plus $f$-regular and $f$-server models in \cite{Chang00}, which are analogous to the arrival curve and service curve models in \cite{NetCal}, try to avoid the implication of packetization by assuming constant packet size, that the service rate is measured in packets per second, and that packets arrive at discrete times (see Sec. 1.1 in \cite{Chang00}). {\em Under these assumptions}, if a server has serving rate $c$ in packets per second, it can be considered as an $f$-server with $f(t)=ct$ or provides a service curve $ct$, see e.g. Example 2.3.2 in \cite{Chang00}. 

However, when packets have variable lengths and the serving rate is not measured in packets per second, special care is needed \cite{Chang00} \cite{NetCal}. To this aim, the max-plus network calculus branch has been motivated \cite{905688}. Specifically, the max-plus $g$-regular and $g$-server models have been introduced to deal with packetization and variable length packets \cite{Chang00} \cite{905688}, which will be exploited in this paper. 

Since the literature NC bounds for service, delay and backlog in TSNs heavily rely on $\beta^{SP*}$ and $\beta^{CBS*}$ as service curves for SP and CBS respectively, the discussion above implies that the proofs of those bounds should be re-examined and the bounds may also need to be updated.

\section{NC-Based Delay Bound Analysis Revisited}\label{sec-5}

In this section, we first examine the delay definitions and their relation in the two branches of network calculus to motivate exploiting the max-plus branch. Then, two approaches are introduced for delay bound analysis. The $(\alpha \to g, \beta \to g)$ approach is to map the min-plus models to the max-plus models and use the max-plus models to perform delay bound analysis. The other approach, called the  $(\alpha, \beta \to g^{x})$ approach,  proposes to combine models from both branches in the analysis. Specifically, it combines the min-plus arrival curve traffic model with the extended max-plus server model, i.e. the $g^{x}$-server model. 

\subsection{Delay, Virtual Delay and Their Relation}

Recall that the delay of a packet $n (\ge 1)$, denoted by $D(n)$, is 
$D(n) = d(n) -a(n)$, which is focused in the max-plus network calculus. 
In the min-plus network calculus, for delay analysis, the focus is on virtual delay 
$D(t) \equiv \inf\{\tau \ge0: A(t) \le A^{*}(t+\tau)\}.$
Consider the tightest upper bounds (if exist) on $D(n)$ and $D(t)$, which are respectively 
\begin{eqnarray}
 \max_{n \ge 1}\{d(n) -a(n)\} &\equiv& D^{(max, +)} \\
\sup_{t \ge 0}\inf\{\tau: A(t) \le A^{*}(t+\tau)\} &\equiv& D^{(min, +)} 
\end{eqnarray}

Proposition \ref{pr-4} proves that an upper bound on virtual delay is also an upper bound on packet delay. An implication is that, directly working on $d(n) -a(n)$ may lead to finding tighter delay bounds. 

\begin{proposition}\label{pr-4}
For any FIFO system without loss, if the delay of any packet is upper-bounded, there holds:
$$
D^{(min, +)} \ge D^{(max, +)}. 
$$
\end{proposition}

\begin{proof}
Consider any packet $n \ge 1$. Note that at $a(n)$, there may be multiple concurrent arrivals. Without loss of generality, suppose $n$ is the last among them and focus on studying its delay, since by FIFO, this packet experiences at least the same or generally higher delay than the other packets among them.  Clearly, $A(a(n)^{+}) = \sum_{k=1}^{n}l(k)$. By definition of $D^{(min, +)}$, we have 
$$
\sum_{k=1}^{n}l(k) = A(a(n)^{+}) \le A^{*}(a(n)^{+}+D^{(min, +)})
$$
Hence 
$$
a(n)^{+}+D^{min, +} \ge {A^{*}}^{\downarrow}(\sum_{k=1}^{n}l(k))
$$
In addition, we must have ${A^{*}}^{\downarrow}(\sum_{k=1}^{n}l(k)) > d(n)$ because $A^{*}(t)$ reaches $\sum_{k=1}^{n}l(k)$ only after packet $n$ has finished its service. Consequently, we have 
$
a(n)^{+}+{D}^{(min, +)} > d(n) 
$
and ${D}^{(min, +)} > d(n) - a(n)^{+}$, or by letting $\epsilon \to 0$, we have ${D}^{(min, +)} \ge d(n) - a(n)$. Since this inequality holds for all $n \ge 1$, we have proved
$$
D^{(min, +)} \ge \max_{n} \{d(n) - a(n)\} = D^{(max, +)}
$$
\end{proof}

\subsection{The $(\alpha \to g, \beta \to g)$ Approach}

In this approach, the min-plus traffic and service models are made direct use of. By mapping them to their max-plus counterparts, improvement on the delay bound results may be immediately found. 

The mappings between the min-plus and max-plus models are summarized in Lemma \ref{lm-map}. Similar mappings have been introduced in the literature, e.g. Lemma 6.2.8 in \cite{Chang00}, Corollary 11.1 and Corollary 11.3 in \cite{JL17}, and Proposition 1 in \cite{Ehsan23}, under their  settings, e.g. discrete time domain and constant packet size when min-plus models are used in \cite{Chang00}, and ``fluid flow arrival functions'' to establish a duality between the two network calculus branches in \cite{JL17}.  For completeness, the proof is included in the Appendix. 

\begin{lemma}\label{lm-map}
(i.a) If a flow has an arrival curve $\alpha$, it is $g_1$-regular with $g_1(v) =\alpha^{\downarrow}(v+l^{m})$.  
(i.b) Conversely if a flow is $g_1$-regular, it has an arrival curve $\alpha(t) = g_1^{\uparrow}(t) + l^{M}$.

(ii.a) If a system provides a service curve $\beta(t)$, it is a max-plus $g_2$-server with $g_2(v) =\beta^{\uparrow}(v)$. 
(ii.b) Conversely, a $g_2$-server provides a service curve $\beta(t) = g_2^{\downarrow}(t)$. 
\end{lemma}

With Lemma \ref{lm-map}, Theorem \ref{th-1} follows immediately from the max-plus delay bound $D^{(g, g)}$ shown in (\ref{gg-db2}). 

\begin{theorem}\label{th-1}
If the input to a system has an arrival curve $\alpha$ and the system provides to the input a service curve $\beta$ or is a $g$-server with $g=\beta^{\uparrow}$, then, for any $n$, its delay is upper-bounded by 
\begin{equation}\label{map-db3}
 V(\alpha^{\downarrow}(v+l^{m}), \beta^{\uparrow}(v)) \equiv D^{(\alpha \to g, \beta \to g)} 
\end{equation}
\end{theorem}

In particular, if the arrival curve is of the token-bucket type and the service curve of the latency-rate type, the following corollary is obtained. 

\begin{corollary}\label{cor-1}
Suppose the input is token-bucket $(\sigma, \rho)$-constrained with arrival curve $\alpha(t) = \rho t + \sigma$, and the service has a latency-rate service curve $\beta(t) = R(t-T)^+$ or is a $g$-server with $g(v)=\frac{v}{R}+T$. If $\rho \le R$, the delay of any packet $n (\ge 1)$ is upper-bounded by  
$\frac{\sigma - l^{m}}{R} + T.$
\end{corollary}

As a specific example, consider a queue exclusively served by a link with rate $c$. It is known that the link provides a latency-rate service curve $c(t-\frac{l^{M}}{c})^{+}$\cite{Jiang03}. Then, the delay bound can be further written as:
\begin{equation}
D^{\alpha \to g, \beta \to g} = \frac{\sigma}{c}  + \frac{l^{M}-l^{m}}{c}.
\end{equation}
As a comparison, if the packetization effect were ignored and $ct$ were used as the service curve, the delay bound, which can be directly found from  $D^{\alpha, \beta}$ in (\ref{ab-db1}), would have been $\frac{\sigma}{c}$ which is clearly better than $D^{\alpha \to g, \beta \to g}$. This implies that such bounds ignoring the packetization effect remain to be verified. Notice also that as implied by Theorem \ref{th-1}, while using the max-plus traffic model helps tighten the delay bound, the effect of changing the service model from min-plus service curve to max-plus $g$-server does not show immediate effect.

\subsection{$g^{x}$-Server and the $(\alpha, g^{x})$ Approach} 
In this subsection, we propose an extension to the max-plus $g$-server model, called $g^{x}$-server, and use it with the min-plus arrival curve model to prove new delay bound results. We call this approach the $(\alpha, g^{x})$ approach.  

\begin{definition}\label{def-gx}
A system is said to be a $g^{x}$-server, with functions $g, x \in \mathcal{F}$, to the input $a(n)$, $n=1, \dots$, if for the output, there holds for all $n \ge 1$:
\begin{equation}\label{max-sc}
d(n) \le \max_{0 \le m \le n} \{a(m) + g(L(m,n)\} +x(l(n)). 
\end{equation}
The $g^{x}$-server is said to be exact, if (\ref{max-sc}) is equation, i.e. $d(n) = \max_{0 \le m \le n} \{a(m) + g(L(m,n)\} +x(l(n))$. 
\end{definition}

It is easily seen that by letting $x(v)=0$, the $g^{x}$-server definition becomes the $g$-server definition. In other words, the $g$-server model is a special case of $g^{x}$-server with $x(v)=0$. In addition, since we have $\max_{0 \le m \le n} \{a(m) + g(L(m, n))\} \le \max_{0 \le m \le n} \{a(m) + g(L(m,n))-\inf_n x(l(n)) \} +x(l(n))$ and $\max_{0 \le m \le n} \{a(m) + g(L(m,n))\} +x(l(n)) \le \max_{0 \le m \le n} \{a(m) + g(L(n)-L(m))+\sup_n x(l(n)) \} $, and since $x(v)$ is non-decreasing, the mappings in Proposition \ref{pr-5} can be verified from the definitions. With Proposition \ref{pr-5}, the mappings between the $g^{x}$-server model and the min-plus service curve model can further be established from Lemma \ref{lm-map}. 

\begin{proposition}\label{pr-5}
(i) A $g^{x}$-server with functions $g$ and $x$ is a $g_1$-server, and (ii) conversely, a $g$-server is a $g_2^{x}$-server with a chosen function $x \in \mathcal{F}$ and an accordingly calculated function $g_2$ , where, 
\begin{eqnarray} 
g_1(v) & = & g(v)+x(l^{M}) \nonumber\\ 
g_2(v) & = & g(v)-x(l^{m}) \nonumber  
\end{eqnarray}
\end{proposition}

As another example, consider the Guaranteed Rate (GR) server model \cite{GR}, which can be used to characterize a wide range of scheduling algorithms \cite{GR} \cite{Jiang03}. A system is said to be a GR server with rate parameter $R$ and error term parameter $E$, if for any packet $n (\ge 1)$, its departure time satisfies:
$$
d(n) \le GRC(n) + E
$$
where $GRC(n)$ can be written as
$$
GRC(n)= \max_{0 \le m \le n} \{a(m) + \frac{\sum_{k=m}^{n-1}l(k)}{R}\} + \frac{l(n)}{R}. 
$$
Comparing the GR definition with the $g^{x}$-server definition, it can be verified that the GR server model is also a special case of $g^{x}$-server, with $g(v)=\frac{v}{R}+E$ and $x(v)=\frac{v}{R}$. 

\vspace{6pt}
Theorem \ref{th-2} presents the delay bound under the $(\alpha, g^{x})$ approach.

\begin{theorem}\label{th-2}
Suppose the arrival has an arrival curve $\alpha$ and the system is a $g^{x}$-server satisfying $x(w) \le g(v+w)-g(v)$. For any packet $n \ge 1$, its delay is upper-bounded by 
\begin{equation}\label{int-db4}
V(\alpha^{\downarrow}, g) 
\equiv D^{(\alpha, g^{x})}
 \end{equation}
\end{theorem}

\begin{proof}
Consider any packet $n (\ge 1)$. The arrival curve definition tells that for any $0 \le m \le n$, $A(a(n)^+) - A(a(m)) \le \alpha(a(n)^+ -a(m))$. Because multiple packets may arrive at $a(n)$ and $A(a(m))$ only includes packets up to $m-1$, we have $A(a(m), a^+(n)) \ge L(n)-L(m)+l(n) = L(m, n+1)$. Hence, there holds 
$$
L(m, n+1) \le A(a(m), a^+(n)) \le \alpha(a(n)^{+} -a(m))
$$  
Taking the lower inverse yields
$$
a(n) -a(m) + \epsilon \ge \alpha^{\downarrow}(L(n+1)-L(m)) 
$$

In addition, since the system is a $g^x$-server with $x(l(n)) \le g(v+l(n))-g(v)$, we have
\begin{eqnarray}
&& d(n) - a(n) \nonumber\\
&\le& \max_{0 \le m \le n} \{a(m) - a(n) + g(L(n)-L(m))\} +x(l(n)) \nonumber\\
&\le&  \max_{0 \le m \le n} \{a(m) - a(n) + g(L(n)-L(m) + l(n))\} \nonumber\\
&=& \max_{0 \le m \le n} \{a(m) - a(n) + g(L(m, n+1))\} \nonumber\\
&\le& \max_{0 \le m \le n} \{g(L(m, n+1)) - \alpha^{\downarrow}(L(m, n+1)) + \epsilon\}  \nonumber\\
&\le& \sup_{v \ge 0} \{g(v) - \alpha^{\downarrow}(v) + \epsilon\}  \nonumber
\end{eqnarray}
Letting $\epsilon \to 0$ completes the proof. 
\end{proof}

In particular, if the arrival curve is of the token-bucket type and the service is a $g^x$-server of GR type, the following corollary is obtained.

\begin{corollary}\label{cor-2}
Suppose the input has arrival curve $\alpha(t) = \rho t + \sigma$, and the system is a $g^x$-server with $g(v) =\frac{v}{R}+E$ and $x(v) = \frac{v}{R}$. 
If $\rho \le R$, the delay of any packet $n (\ge 1)$ is upper-bounded by 
$\frac{\sigma}{R} + E . $
\end{corollary}
\begin{proof}
The lower inverse function of  $\alpha(t)$ is: 
\[
    \alpha^{\downarrow} (v) = 
\begin{cases}
    \frac{v-\sigma}{\rho},& \text{if } v\geq \sigma\\
    0,              & \text{otherwise}
\end{cases}
\]
Hence
\[
    \frac{v}{R} - \alpha^{\downarrow} (v)  = 
\begin{cases}
    \frac{v}{R} - \frac{v-\sigma}{\rho} = \frac{\sigma}{R} + (\frac{v-\sigma}{R}- \frac{v-\sigma}{\rho}),& \text{if } v\geq \sigma\\
    \frac{v}{R},              & \text{otherwise}
\end{cases}
\]
where, for $\rho \le R $, $\frac{v-\sigma}{R}- \frac{v-\sigma}{\rho} \le 0$ and hence the supremum is $\frac{v}{R}$ in both cases. Since it is a $g^x$-server with $g(v)=\frac{v}{R}+E$, we have the delay bound from Theorem \ref{th-2}:
$$
\sup_{v \ge 0} \{ \frac{v}{R}+E - \alpha^{\downarrow} (v) \} = \frac{\sigma}{R} + E. 
$$
\end{proof}

As a specific example, consider again the case that a queue is exclusively served by a link with rate $c$. It is also known that the system is a GR server to the queue, with $R=c$ and $E=0$ \cite{Chang00}\cite{Jiang03}. The delay bound can be further written as:
\begin{equation}
D^{(\alpha, g^x)} = \frac{\sigma}{c}
\end{equation}
which is the same as and hence validates the min-plus bound obtained by treating as if the link would provide a service curve of $c t$, ignoring the packetization effect. 

To check the tightness of the bound $D^{(\alpha, g^{x})}$, consider the above case with the traffic from an ATS shaper that emulates token bucket \cite{8021Qcr}. Suppose there is enough traffic to be processed and the bucket is full at some time instance. Then, up to $\sigma$ amount of traffic will be sent out from the shaper instantaneously. Consider the last packet in the burst, which will experience the longest delay among packets in the burst, and it can be verified that the delay for its last bit to finish transmission on the link and hence its delay is $\frac{\sigma}{c}$ that equals $D^{(\alpha, g^x)} $. With this, we can conclude: 

\begin{proposition}\label{pr-6} 
The bound $D^{\alpha, g^{x}}$ can be reached. 
\end{proposition}

\subsection{Comparison}
In this and the previous sections, four approaches have been introduced for delay bound analysis, which are the min-plus-only approach with delay bound as $D^{(\alpha, \beta)}$ in (\ref{ab-db1}), the max-plus-only approach with delay bound as $D^{(g, g)}$ in (\ref{gg-db2}), the min-plus to max-plus mapping approach with delay bound as $D^{(\alpha \to g, \beta \to g)}$ in (\ref{map-db3}), and the integrated approach with delay bound as $D^{(\alpha, g^{x})}$ in (\ref{int-db4}). 

Consider again the single link case, which is a queue exclusively served by a link with rate $c$ and the input has a token-bucket arrival curve $\rho t + \sigma$ with $\rho \le c$. As discussed previously in this section, the link has a service curve $c(t-\frac{l^M}{c})^{+}$ and is a $g$-server with $g(v)=\frac{v}{c}+\frac{l^M}{c}$. Applying these, different delay bounds can be computed from the four approaches. Table~\ref{tb2} presents and compares these bounds. As is clear from the table, $D^{(\alpha, g^{x})}$ is the tightest. Indeed, as the discussion preceding Proposition \ref{pr-6} shows, the delay bound $\frac{\sigma}{c}$ by $D^{(\alpha, g^{x})}$ may be reached. 

\begin{table}[htb]
\caption{Delay bound comparison} \label{tb2}
\centering
\begin{tabular}{|r|l|}
\hline
$D^{(\alpha, \beta)}$: (\ref{ab-db1}) & $\frac{\sigma}{c}+\frac{l^M}{c}$ \\ \hline
$D^{((\alpha \to) g, g)}$: (\ref{gg-db2}) & $\frac{\sigma}{c}+\frac{l^M}{c}-\frac{l^{m}}{c}$ \\ \hline
$D^{(\alpha \to g, \beta \to g)}$: (\ref{map-db3}) & $\frac{\sigma}{c}+\frac{l^{M}-l^{m}}{c}$ \\ \hline
$D^{(\alpha, g^{x})}$: (\ref{int-db4}) & $\frac{\sigma}{c}$ \\ \hline
\end{tabular}%
\end{table}

As a remark, another related approach is adopted in \cite{Ehsan23} to improve NC delay bounds, where the min-plus arrival curve model is mapped to the max-plus $g$-regular traffic model, while the min-plus service curve model is still used. As implied by part (ii) in Lemma \ref{lm-map}, this approach can give the same delay bound as the $D^{\alpha \to g, \beta \to g}$ approach. 

\section{Service and Delay Bounds for SP and CBS in TSNs} \label{sec-6}

In this section, the $(\alpha, g^x)$ approach, motivated by Proposition \ref{pr-6}, is utilized to obtain service and delay bounds for TSNs. The focused transmission selection algorithms are SP and CBS. We first study them as standalone systems to gain insights, and then consider systems where both are used.  
Specifically, we consider a queue that may share with other queues for transmission over a link with rate $c$. The transmission selection among queues uses SP. The traffic to each queue $i (\ge 1)$ is token bucket $(\sigma_i, \rho_i)$-constrained. Without loss of generality, a queue with a smaller index number has a higher priority, implying queue 1 is given the highest priority. 

\subsection{Standalone SP and CBS} 
\subsubsection{SP}
When SP is used alone, Theorem \ref{th-sp} presents bounds on the service and delay to each queue in the system, where, for the considered queue at priority level $i$, $\rho_u = \sum_{j =1}^{i-1} \rho_{j}$, $\sigma_u = \sum_{j =1}^{i-1}\sigma_{j}$, $l^{m_i}$ denotes the minimum packet length of queue $i$, and $l^{M_l}$ denotes the maximum packet length of lower priority queues than $i$. The proof is included in the Appendix. 

\begin{theorem}\label{th-sp}
(i) The service provided to a queue $i$ is a $g^{x}$-server with 
\begin{eqnarray}
g(v) &=& \frac{v}{c-\rho_u} + E \\
x(v) &=& \frac{v}{c-\rho_u} 
\end{eqnarray}
where
\begin{eqnarray}
E &=& \frac{\sigma_u + l^{M_l}}{c-\rho_u} - \frac{l^{m_i}}{c-\rho_u}+ \frac{l^{m_i}}{c}. \nonumber
\end{eqnarray}
 (ii) If $\rho_i \le c-\rho_u$, the delay of any packet to the queue is upper-bounded by 
\begin{equation}\label{sp-db1}
 \frac{\sigma_i}{c - \rho_u} + \frac{\sigma_u + l^{M_l}}{c-\rho_u} - \frac{l^{m_i}}{c-\rho_u}+ \frac{l^{m_i}}{c}
\end{equation}
\end{theorem}

Let $l^{M}$ denote the maximum packet length in all queues. As a comparison, in the LRQ work \cite{ECRTS16}, using a timing analysis method, the following bound has been found: 
\begin{equation}\label{sp-db2}
\frac{\sigma_i}{c - \rho_u} +  \frac{\sigma_u + l^{M_l} }{c-\rho_u}+  \frac{l^{M} }{c}. 
\end{equation}
In addition, the following delay bound is obtained by using the service curve model that has ignored the packetization effect (cf. \cite{Zhao22} and references therein):
\begin{equation}\label{sp-db3}
\frac{\sigma_f}{c - \rho_u} +  \frac{\sigma_u + l^{M_l} }{c-\rho_u}+  \frac{l^{M} }{c-\rho_u}.
\end{equation}
Clearly, the timing-analysis based bound (\ref{sp-db2}) is better than (\ref{sp-db3}).
The difference is $\frac{l^{M} }{c-\rho_u} - \frac{l^{M} }{c}$. Recently in \cite{Ehsan18} and \cite{Ehsan23}, there is an effort to improve the delay bound (\ref{sp-db3}), which also results in (\ref{sp-db2}). Recall the discussion in Section \ref{sec-4}:  When proving the service curve analysis based bound (\ref{sp-db3}), the literature has used service curves overlooking the packetization effect. 
The bound (\ref{sp-db1}) from the $(\alpha, g^x)$ approach makes further improvement with a reduction of 
 $\frac{l^{M} - l^{m_i}}{c} + \frac{l^{m_i}}{c-\rho_u}$ from (\ref{sp-db2}), implying also a validation and proof of (\ref{sp-db3}). This is similar to the comparison in Table \ref{tb2}.

In the proof of Theorem \ref{th-sp}, we have shown $d(n) \le \max_{0\le m \le n}\{ a(m)+ \frac{L(m,n)}{c-\rho_u} +E \} + \frac{ l(n)}{c-\rho_u} $ from which it is easily verified $d(n) \le \max_{0\le m \le n}\{ a(m)+ \frac{L(m,n)}{c-\rho_u} +E + \frac{ l^{M_i}}{c-\rho_u} \} $. Then, Corollary \ref{cor-3} follows from the definition of $g$-server and part (ii.b) of Lemma \ref{lm-map}. 

\begin{corollary} \label{cor-3}
The service provided to a queue $i$ is a $g$-server with $g(v) = \frac{v+l^{M_i}}{c-\rho_u} +E$ and provides a service curve $\beta(t)= (c-\rho_u) (t- T)^{+}$ with $T=E+\frac{ l^{M_i}}{c-\rho_u}$, where $l^{M_i}$ denotes the maximum packet length of the queue.
\end{corollary}

As a special case, consider the link with rate $c$. By removing factors due to traffic from other queues, i.e. letting $0=\rho_u=\sigma_u=l^{M_l}$, the same service curve $c(t-\frac{l^{M_i}}{c})^{+}$ for the link is recovered from Corollary \ref{cor-3}. In addition, the same delay bound $\frac{\sigma_i}{c}$ is recovered from Theorem \ref{th-sp}. 

\vspace{6pt}
\subsubsection{CBS}
Consider a standalone credit-based shaper with $idleSlope$, whose queue transmits packets on a link with rate $c$. We also have service and delay bounds as shown in Theorem \ref{th-cbs}, where $I \equiv idleSlope$, $S \equiv sendSlope$ and $S=I-c$.  

\begin{theorem} \label{th-cbs}
(i) The CBS shaper is an exact $g_1^{x}$-server with $g_1(v) = \frac{v}{I}$ and $x(v) = \frac{v}{c}$; and a $g_{2}^{x}$-server with $g_{2}(v) = (\frac{v}{I} + E)^{+}$ and $x(v)= \frac{v}{I}$ where $E=(\frac{1}{c} - \frac{1}{I})l^{m}$.

(ii) If the input is token-bucket $(\sigma, \rho)$ constrained and $\rho \le I$, the delay of any packet is upper-bounded by
\begin{eqnarray}
 \frac{\sigma}{I} + (\frac{1}{c} - \frac{1}{I})l^{m}
\end{eqnarray}
\end{theorem}

\begin{proof}

\begin{figure}[ht!]
\centering
  \includegraphics[width=0.8\linewidth]{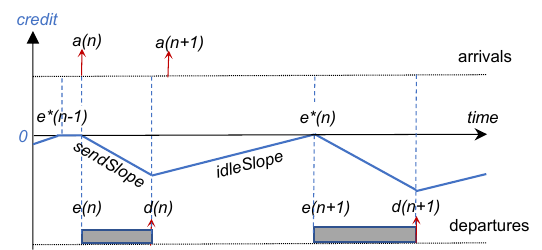} 
  \caption{$Credit$ function in CBS without conflicting traffic} 
  \label{fig-cbs}
\end{figure}

For (i), consider any packet $n (\ge 1)$. 
Let $e(n)$ denote the time when packet $n$ enters transmission, and {\em $e^*(n)$ the moment at which the increase of $credit$ due to the transmission of packet $n$ ends}, with $e^*(0)$ set to $0$. The relation between $e(n)$ and $e^*(n)$ is, as also indicated by Figure \ref{fig-cbs}, 
\begin{equation} \label{eq-tm1} 
    e(n)= 
\begin{cases}
    e^*(n-1),& \text{if } a(n)\leq e^*(n-1)\\
    a(n),              & \text{if } a(n) > e^*(n-1)
\end{cases}
\end{equation}
which is, 
\begin{equation}  \label{eq-tm1a}
   e(n)= \max \{a(n), e^*(n-1)\}.
\end{equation}
In addition, we have $d(n) - e(n) = \frac{l(n)}{c}$, from which the value of $credit$ at $d(n)$ can be calculated as $sendSlope (d(n)-e(n))$, with which, it can be further calculate:
$e^*(n) - d(n) = - credit / idleSlope = \frac{-sendSlope}{idleSlope}(d(n)-e(n))$. With these, it can be verified: 
\begin{eqnarray}
e^*(n) - e(n) &=&  \frac{l(n)}{idleSlope} \label{eq-tm2}
\end{eqnarray}

Note that the above discussion is valid for any $n$, so we also have $e^*(n-1)  = e(n-1) + \frac{l(n-1)}{idleSlope} $. Applying to (\ref{eq-tm1a}), we have
\begin{eqnarray}
e(n) &=& \max \{a(n), e^*(n-1)\}  \nonumber\\
&=& \max \{a(n), e(n-1) + \frac{l(n-1)}{idleSlope}\} 
\end{eqnarray}
Applying the right hand side iteratively leads to
\begin{eqnarray}
e(n) &=& \max \{a(n), e^*(n-1)\}  \nonumber\\
&=& \max_{m=0}^{n} \{a(m) + \sum_{k=m}^{n-1} \frac{l(m)}{idleSlope}\}  \nonumber\\
&=& \max_{m=0}^{n} \{a(m) + \frac{L(n)-L(m)}{idleSlope}\} 
\end{eqnarray}
 Since $d(n) = e(n) + \frac{l(n)}{c}$, there holds
\begin{eqnarray}
d(n) &=& \max_{m=0}^{n} \{a(m) + \frac{L(m,n)}{I}\}  + \frac{l(n)}{c} \label{eq-cbs1} \\
&=& \max_{m=0}^{n} \{a(m) + \frac{L(m,n)}{I}+ \frac{l(n)}{c}  - \frac{l(n)}{I} \}  + \frac{l(n)}{I} \nonumber\\
&\le& \max_{m=0}^{n} \{a(m) + \frac{L(m,n)}{I} + (\frac{1}{c} - \frac{1}{I})l^{m}\} + \frac{l(n)}{I} \nonumber\\
\label{eq-cbs3}
\end{eqnarray}
The exact $g_1^{x}$-server part is proved by (\ref{eq-cbs1}) and $g_2^{x}$-server part (i.b) is proved by (\ref{eq-cbs3}). 

With $g_2$-server representation in (i), part (ii) follows immediately from Corollary \ref{cor-2}.
\end{proof}

From (\ref{eq-cbs1}) in the proof of Theorem \ref{th-cbs}, we can further derive for any $n (\ge 1)$, $d(n) \le \max_{m=0}^{n} \{a(m) + \frac{L(m,n)}{idleSlope}+ \frac{l^{M}}{c}\}$, which together with part (ii.b) of Lemma \ref{lm-map} leads to Corollary \ref{cor-4}. In contrast to the literature, $\frac{l^{M}}{c}$ is included in the service curve, factoring in the packetization effect. 

\begin{corollary} \label{cor-4}
The CBS shaper is a $g$-server with $g(v) = \frac{v}{idleSlope} +\frac{l^{M}}{c}$ and provides a service curve $\beta(t)=idleSlope (t- \frac{l^{M}}{c})^{+}$.
\end{corollary}

\subsection{SP and CBS in Combination}  
For simplicity, we consider two settings where only one of the priority queues applies CBS. Depending on the position of this CBS queue, the CBS credit value may be affected by high priority traffic. 
These settings together with the SP standalone setting are fundamental settings for TNSs \cite{8021Q} \cite{8021BA} and have been focused in the NC based TSN analysis literature, e.g. \cite{MB14, Zhao18, Ehsan18, Zhao21, Zhao22}.

\subsubsection{CBS at the highest priority} 
In this setting, the CBS queue is queue 1, i.e. the queue with the highest priority. The service and delay bounds to this CBS queue are summarized in Theorem \ref{th-cbs2}. 

\begin{theorem}\label{th-cbs2}
(i) For the CBS queue at the highest priority, the system is a $g_1^{x}$-server with  
\begin{eqnarray}
g_1(v) &=& \frac{v}{I} + \frac{l^{M_l}}{c} \nonumber\\ 
x_1(v) &=& \frac{v}{c}
\end{eqnarray}
and a $g_2^{x}$-server with
\begin{eqnarray}
g_2(v) &=& (\frac{v}{I} + E_2)^{+} \nonumber\\
x_2(v) &=& \frac{v}{I}
\end{eqnarray}
with $$E_2=\frac{l^{M_l}}{c} - (\frac{1}{I}-\frac{1}{c})l^{m_1}.$$

(ii) If the traffic to the CBS queue is token-bucket $(\sigma, \rho)$ constrained and $\rho \le I$, the delay of any packet to the CBS queue is upper-bounded by
\begin{eqnarray}
 \frac{\sigma}{I} + \frac{l^{M_1}}{c} - (\frac{1}{I}-\frac{1}{c})l^{m_1} 
\end{eqnarray}
\end{theorem}

\begin{proof}
To assist, Figure \ref{fig-cbs2} uses an example to show how the value of $credit$ progresses over time and when each packet enters (becomes eligible for) transmission and finally departs. Different from Figure \ref{fig-cbs} where there is no conflicting traffic, in Figure \ref{fig-cbs2}, conflicting traffic from lower priority may be in presence. 

\begin{figure}[h!]
\centering
  \includegraphics[width=0.98\linewidth]{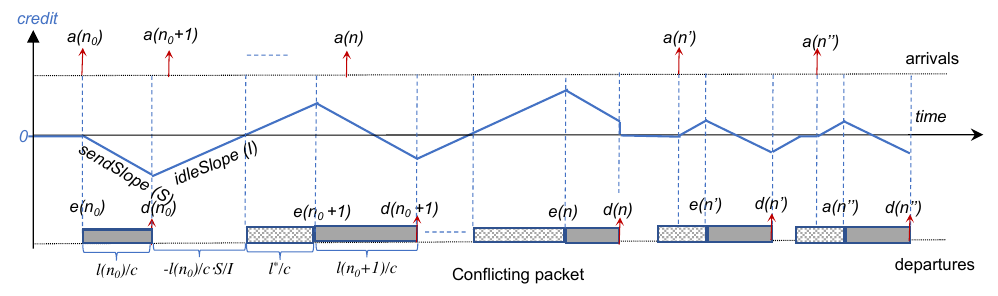} 
  \caption{$Credit$ function in CBS with conflicting traffic from lower priority} 
  \label{fig-cbs2}
\end{figure}

For any packet $n (\ge 1)$ from the considered CBS queue, let $e(n)$ denote the time when the packet enters transmission. 
In addition, let $n_0 ( \le n)$ denote the nearest packet satisfying that immediately before its arrival there is no CBS packet in the system and the value of $credit$ is zero. Note that such an $n_0$ always exists, because packet $1$ is such a packet. In other words, 
$$n_0 = \max\left\{m (\le n): credit (a(m))=0 ; a(m)>d(m-1)\right\}$$ 

Note that, according to CBS, for $n_0 < m \le n$, $a(m) > d(m-1)$ and $credit (a(m))\ge0$ cannot happen together. This is because, when $a(m) > d(m-1)$, implying that the CBS queue is empty, there are two cases. (a) If $credit(d(m-1))$ were still positive after the transmission of packet $m-1$, the value of $credit$ would be set to zero and this zero value would continue till $a(m)$; or (b) if $credit(d(m-1))$ were negative,  $credit$ would increase at rate $idleSlope$ until zero, and with $credit (a(m))\ge0$, we would have had $a(m)$ after $credit$ had reached zero from negative. In both cases (a) and (b), we would have $credit (a(m))=0 \textrm{ and } a(m)>d(m-1)$ and hence $n_0$ would have been this $m$. The two cases are also illustrated in Figure \ref{fig-cbs2}, at the right end.
The discussion implies that there is no credit-holding with $credit (a(m))=0$ in $(a(n_0), d(n))$. 

Let $\Delta t\equiv d(n) - a(n_0) $, and let $\Delta t^{\uparrow}$ (respectively $\Delta t^{\downarrow}$) denote the accumulated length of all periods in $[a(n_0), d(n)]$ when $credit$ is increasing (respectively decreasing). Since there is no holding periods in $\Delta t$, we have 
\begin{equation} \label{eq-dt}
\Delta t = \Delta t^{\uparrow} +  \Delta t^{\downarrow}
\end{equation}

According to the operation of CBS,  $credit$ is decreasing only during one packet's transmission, the length of each $credit$-decreasing period is the transmission time of the corresponding packet $l(m)/c$, and in $[a(n_0), d(n)]$, such packets are $n_0, \dots, n$. Hence, we have
\begin{eqnarray}
\Delta t^{\downarrow} & = & \sum_{m=n_0}^{n}\frac{l(m)}{c} = \frac{L(n+1)-L(n_0)}{c}\label{eq-dt-}
\end{eqnarray}

In addition, the credit decreasing rate is $sendSlope \equiv S$ in periods of $\Delta t^{\downarrow}$, and the increasing rate is $idleSlope \equiv I$ in $ \Delta t^{\uparrow}$, and at the beginning of the period, $credit(a(n_0))=0$, so
\begin{eqnarray}
&& credit(d(n))\nonumber \\ 
& = & credit(a(n_0))+ sendSlope \cdot \Delta t^{\downarrow} + idleSlope \cdot   \Delta t^{\uparrow} \nonumber \\ 
&=& S \cdot \Delta t^{\downarrow}  + I \cdot \Delta t^{\uparrow} \label{eq-credit}
\end{eqnarray}
Moreover, $credit(d(n))=credit(e(n))+ S \frac{l(n)}{c}$ and $credit(e(n)) \ge 0$. 

Let $s_0$ denote the time before $e(n)$ such that $credit$ is just increased from negative to zero at $s_0$. According to the operation of CBS, the increase from $credit(s_0) = 0$ to $credit(e(n))$ happens only when the CBS queue is not empty and the increase is due to transmission of packets not from this CBS queue. Since the CBS queue has the highest priority, at most one packet from lower priorities can contribute to this increase. Hence $credit(e(n)) - credit(s_0) \le I \frac{l^{M_l}}{c}$ or $credit(e(n)) \le I \frac{l^{M_l}}{c}$, with which, we have:
\begin{eqnarray}
credit(d(n))& \le &  I \frac{l^{M_l}}{c} + S \frac{l(n)}{c} \equiv credit^{M}\label{eq-credit-n}
\end{eqnarray}
Applying to (\ref{eq-credit}), we obtain $$S \cdot \Delta t^{\downarrow}  + I \cdot \Delta t^{\uparrow} \le  credit^{M}$$ and hence
\begin{eqnarray}
\Delta t^{\uparrow} &\le& \frac{- S }{I}  \Delta t^{\downarrow} + \frac{credit^{M} }{I}  \label{eq-dt+}
\end{eqnarray}

\nop{
Moreover, it can be shown, e.g. Lemma 1 in \cite{MB14}, that in this case, $credit$ is bounded: $credit^{m} \le credit \le credit^{M}$, 
with $credit^{M} = \frac{I \cdot l^{M_l}}{c}$ and $credit^{m} = \frac{S \cdot l^{M_1}}{c}$, 
where $l^{M_l}$ denotes the maximum packet length of conflicting traffic, and $l^{M_1}$ that of the CBS traffic. Applying to (\ref{eq-credit}), we obtain $$S \cdot \Delta t^{\downarrow}  + I \cdot \Delta t^{\uparrow} \le  credit^{M}$$ and hence
\begin{eqnarray}
\Delta t^{\uparrow} &\le& \frac{- S }{I}  \Delta t^{\downarrow} + \frac{credit^{M} }{I}  \label{eq-dt+}
\end{eqnarray}
}

Since by the default setting $S=I-c$, $I-S=c$ and the following is resulted by applying (\ref{eq-dt+}) and (\ref{eq-dt-}) to (\ref{eq-dt}): 
\begin{eqnarray}
\Delta t &\le& \Delta t^{\downarrow} + \frac{- S }{I}  \Delta t^{\downarrow} + \frac{credit^{M} }{I} \nonumber\\
&=& \frac{L(n+1)-L(n_0)}{I} + \frac{credit^{M} }{I}
\end{eqnarray}
Since $\Delta t = d(n) - a(n_0)$, we have 
\begin{eqnarray}
&& d(n) \nonumber\\
&\le&  a(n_0) +  \frac{L(n+1)-L(n_0))}{I} +  \frac{credit^{M} }{I} \label{eq-cM}\\
&\le& \max_{0\le k \le n}\{ a(k) +  \frac{L(n)-L(k)}{I} +  \frac{l^{M_l} }{c} \} + \frac{l(n)}{c} \label{eq-cM1} \\
&\le& \max_{0\le k \le n}\{ a(k) +  \frac{L(n)-L(k)}{I} +  E_2 \} + \frac{l(n)}{I} \label{eq-cM2} 
\end{eqnarray}
where in step (\ref{eq-cM1}) we have applied (\ref{eq-credit-n}), i.e. $credit^{M}= I \frac{l^{M_l}}{c} + S \frac{l(n)}{c}$ and $S=I-c$. For the $g^{x}_1$-server part of (i), it is from  (\ref{eq-cM1}), while the $g^{x}_2$-server part from (\ref{eq-cM2}). 

With the $g^{x}_2$-server characterization, part (ii) follows immediately from Corollary \ref{cor-2}.
\end{proof}


\subsubsection{CBS not at the highest priority and credit is on hold when there is high priority transmission} 

This setting is similar to the standalone SP setting in the previous subsection, but here one queue implements CBS and there are other queues with higher priority than the CBS queue. In addition, when a higher priority packet is selected for transmission, the $credit$ counter of the CBS is frozen, i.e. its value is kept unchanged during this high priority packet's transmission, similar to CBS combined with timed gate operation in ETS\cite{8021Q}\cite{8021BA}. 
So, the setting is a way to model a SP+CBS setting where enhancements for scheduled traffic (ETS) are also supported \cite{8021Q}\cite{8021BA}. For a similar setting, a delay bound is introduced in \cite{Ehsan18}, where however the packetization effect on CBS is not considered in its related service curve model. In Theorem \ref{th-cbs4}, we prove an improved delay bound based on the proposed $(\alpha, g^{x})$-approach, which further validates the bound in \cite{Ehsan18}. 

\begin{theorem}\label{th-cbs4}
(i) For the CBS queue $i$, the system is a $g_1^{x}$-server to it with 
\begin{eqnarray}
g_1(v) &=&  \frac{v}{R}  + \frac{\sigma_{u} + l^{M_l}}{c- \rho_{u}} \nonumber\\  
x_1(v) &=&  \frac{v}{c}
\end{eqnarray}
and a $g_2^{x}$-server with 
\begin{eqnarray}
g_2(v) &=&  (\frac{v}{R}  + E_2)^{+} \nonumber\\  
x_2(v) &=&  \frac{v}{R}
\end{eqnarray}
where 
\begin{eqnarray}
R &=&  I\frac{c- \rho_{u}}{c} \nonumber\\  
E_2 &=&  \frac{\sigma_{u} + l^{M_l}}{c- \rho_{u}}  - (\frac{1}{R}-\frac{1}{c})l^{m_i}
\end{eqnarray}

(ii) If the traffic to the CBS queue is token-bucket $(\sigma, \rho)$ constrained and $\rho \le R$, the delay of any packet to the CBS queue is upper-bounded by
\begin{eqnarray}
 \frac{\sigma}{R} + \frac{\sigma_{u} + l^{M_l}}{c- \rho_{u}}  - (\frac{1}{R}-\frac{1}{c})l^{m_i} 
\end{eqnarray}
\end{theorem}

As a cross-check, by removing the higher priority traffic, the results in Theorem \ref{th-cbs2} are recovered from Theorem \ref{th-cbs4}. The proof of Theorem \ref{th-cbs4} is similar to that of Theorem \ref{th-cbs2} and is included in the Appendix. 

\section{Concluding Remarks} \label{sec-7}

For performance guarantees in time-sensitive networks (TSNs), the network calculus theory (NC), particularly its min-plus branch, has been extensively applied to compute worst-case delay bounds. In this paper, a revisit to some of the most fundamental results has been conducted. Specifically, it is shown that the commonly used min-plus service curve models for strict priority (SP) and credit-based shaper (CBS), which are two basic transmission selection algorithms for TSNs, have overlooked the packetization effect. To address, we have examined the delay definition difference and its impact on delay bound analysis, based on which, two approaches are introduced for service and delay bound analysis. 

One approach is to use the max-plus network calculus models as the basis and map the min-plus models to them. The other builds the analysis on extending the max-plus $g$-server model to the $g^{x}$-server and using the extended model with the min-plus traffic model together. In addition, the integrated approach is applied to several SP and CBS settings, for which, service and delay bounds are derived. These bounds not only show general improvement over but also imply the validity of the existing bounds even though they have been computed using the min-plus service curve models where the packetization effect may have been overlooked. 


\bibliographystyle{unsrt}
\bibliography{nc-qt}

\appendix

\noindent{\bf A. Proof of Lemma \ref{lm-map}} 

\begin{proof}
For (i.a), consider any packet $n$. The arrival curve definition tells that for any $0 \le m \le n$, $A(a(n)^+) - A(a(m)) \le \alpha(a(n)^+ -a(m))$. Since multiple packets may arrive at $a(n)$ and $n$ is one of them, we have $A(a(m), a^+(n)) \ge \sum_{k=1}^{n}l(k) = L(n+1)-L(m)$. Hence,  
$$
L(m,n)+l(n) \le A(a(m), a^+(n)) \le \alpha(a(n)^+ -a(m))
$$  
and then 
$$
L(m,n) +  l^{m}  \le \alpha(a(n)^+ -a(m))
$$  
Taking the lower pseudo-inverse yields
$$
\alpha^{\downarrow}(L(m,n) + l^{m} ) \le a(n) -a(m) + \epsilon 
$$
and hence 
$$
a(n)  \ge a(m) + \alpha^{\downarrow}(L(m,n) + l^{m}) - \epsilon
$$
Since the above is proved for any $m, (0 \le m \le n)$, we have 
$a(n)  \ge \max_{0 \le m \le n}\{a(m) + \alpha^{\downarrow}(L(m,n) + l^{m})\} - \epsilon$
and part (i.a) is proved by letting $\epsilon \to 0$. 

For (i.b), consider any time $t >0$. Let $n$ be the first packet that arrives after $t$, i.e. $n = \min\{k: a(k) > t\} $. Hence, $a(n-1) \le t < a(n)$. For any time $(t \ge) s >0$, it is similarly found $a(m-1) \le s < a(m)$ for some $m$. Then, we have $a(n-1) - a(m) \le t-s \le a(n) - a(m-1)$, $A(t) = \sum_{k=1}^{n-1}l(k) = L(n)$ and $A(s) = L(m)$. 
If $m=n$, clearly $A(s,t)=0$. We now consider $m <n$. Since the flow is $g_1$-regular, $a(n-1) - a(m) \ge g_1(L(m, n-1)))$. Taking the upper pseudo-inverse gives $g_1^{\uparrow}(a(n-1) - a(m)) \ge L(m, n-1))$, with which we have 
\begin{eqnarray}
A(s,t) &=& L(n) - L(m) \le  l^{M}+L(m, n-1) \nonumber\\
&\le& l^{M}+ g_1^{\uparrow}(a(n-1) - a(m)) \nonumber
\end{eqnarray}
Since $g^{\uparrow}$ is non-decreasing and $a(n-1) - a(m) \le t-s$, we have
$$
A(s,t) \le l^{M}+ g_1^{\uparrow}(t - s) 
$$
which, together with the case $m=n$, concludes the proof. 

For (ii.a), consider any packet $n \ge 1$ and focus on its departure time $d(n)$. Note that by definition, $A^{*}(t)$ represents the amount of traffic up to time $t$ (excluded), so $A^{*}(d(n))$ does not include the packe(s) finishing at $t$, among which at least one is packet $n$. Hence, $A^{*}(d(n))) \le L(n)$. 

The service curve definition tells that there exists some time $s$, $(0 \le s \le d(n))$, such that $A^{*}(d(n)) \ge A(s) + \beta(d(n) -s)$. 
Let $m=\min\{k: a(k-1) < s\}$. Note that such an $m$ exists. This is because $a(0)=0$ and $a(-1) < 0$ by definition, and hence we always have $k=0$ to ensure $a(k-1) < s$. 

Since $m=\min\{k: a(k-1) < s\}$, we must have $a(m-1) < s \le a(m)$. Here, the first part $a(m-1) < s$ implies that by $ s$ (excluded), at least packets $1, \dots, m-1$ have arrived, and hence $A(s) \ge L(m)$. The second part $s \le a(m)$ together with that $\beta$ is non-decreasing gives $\beta(d(n) -s) \ge \beta(d(n) -a(m))$. Together with $A^{*}(d(n)))\le L(n)$, we now have:  
\begin{eqnarray}
L(n) &\ge& A^{*}(d(n))  \nonumber\\
&\ge& A(s) + \beta(d(n) -s) \nonumber\\
&\ge& L(m) + \beta(d(n) - a(m)) \nonumber
\end{eqnarray}
which gives
$$
L(n) - L(m) \ge \beta(d(n)-a(m)) 
$$
Taking the upper inverse yields
$$
d(n) -a(m) \le \beta^{\uparrow}(L(n)-L(m)) 
$$
and hence, 
\begin{eqnarray}
d(n) &\le& a(m) + \beta^{\uparrow}(L(n)-L(m)) \nonumber\\
&\le& \max_{0 \le m \le n}\{ a(m) + \beta^{\uparrow}(L(n)-L(m)) \} \nonumber
\end{eqnarray}
which completes the proof of (ii.a). 

For (ii.b), consider any time $t >0$. Let $n = \min\{k: d(k) > t\} $, and hence $d(n-1) \le t <d(n)$. 
This implies that all packets up to $n$-1 (included) have finished service. Hence $A^{*}(t) =L(n)$. 
Since it is a $g$-server, by definition, there exists some $(0 \le) m (\le n)$ such that  $d(n) \le a(m) + g(L(n)-L(m))$. Since $t <d(n)$, we hence have 
$$
t \le a(m) + g(L(n)-L(m)) = a(m) + g(A^{*}(t)-L(m)). 
$$
Let $s=a(m)$. Since $A(a(m))$ does not include packets that arrive at $a(m)$, one of which is packet $m$. This means, $A(a(m))$ at most includes packets up to $m-1$. Consequently, we have $L(m) \equiv \sum_{k=1}^{m-1}l(k) \ge A(a(m))$, which leads to 
$$
t \le a(m) + g(A^{*}(t)-A(a(m))
$$
or $t - a(m) \le g(A^{*}(t)-A(a(m))$. 
Taking the lower inverse gives
$$
A^{*}(t)-A(a(m)) \ge g^{\downarrow}(t -a(m))
$$
or 
$A^{*}(t) \ge A(s) + g^{\downarrow}(t -s) \ge \inf_{0\le s \le t} \{A(s) + g^{\downarrow}(t -s)\}$
which completes the proof. 
\end{proof}

\vspace{6pt}
\noindent{\bf B. Proof of Theorem \ref{th-sp}} 
\begin{proof}
For part (i), consider any packet $n (\ge 1)$ to the queue $i$. Suppose the departure time $d(n)$, is within the busy period of the system which starts at $s$. Note that such a busy period always exists, since in the worst case, the period is only the service time period of the packet and in this case, $s=a(n)$. 

Since the link has rate $c$ and it is busy with serving between $s$ and $d(n)$, there holds:
\begin{equation}\label{eq-2}
d(n) = s + \frac{\sum_{k=\tilde{n}_0}^{\tilde{n}} l^{k}}{c},
\end{equation}
where $\tilde{n}_0$ denotes the packet whose arrival starts the busy period, and $\tilde{n}_0$ its packet sequence number and $\tilde{n}$ the sequence number of packet $n$ seen at the other end of the link. 

Among packets $\tilde{n}_0, \dots, \tilde{n}$, some are from the considered queue $i$ and the rest the other queues. Let $n_0$ denote the first packet from queue $i$ in the busy period. There holds $a(n_0) \ge s$. 
Equation (\ref{eq-2}) can be re-written as:
\begin{equation}\label{eq-3aa}
d(n) \le s + \frac{\sum_{k=n_0}^{n} l(k)}{c} + \frac{A^{*}_{u}(s, d(n))+A^{*}_{l}(s, d(n))}{c},
\end{equation}
where $A^{*}_{u}(s, d(n))$ and $A^{*}_{l}(s, d(n))$ respectively represent the total length (in bits) of packets from the lower and higher queues transmitted in $(s, d(n))$. 

Since the busy period starts at $s$, this implies that immediately before $s$, the link is idle. In other words, all packets, which arrived before $s$, have been transmitted by $s$. So, we have $A^{*}_{u}(s) = A_{u}(s)$, $A^{*}_i(s) = A_i(s)$, and $A^{*}_{l}(s) = A_{l}(s)$. In addition, due to priority, there is at most one packet from lower priority queues in $A^{*}_{u}(s) + A^{*}_{l}(s)$, and if there is, it must be the packet that starts the busy period. Moreover, all packets from higher priority queues, which are served before $d(n)$, must have arrived by $d(n)-\frac{l(n)}{c}$. \footnote{This is due to being non-preemptive, so among packets from higher priority queues, only those that have arrived before packet $n$ enters service at $d(n)-\frac{l(n)}{c}$ can be served before $n$. } 
So, we have $A^{*}_{u}(d(n)) \le A_{u}(d(n)-\frac{l(n)}{c})$. Combing these, we obtain:
\begin{eqnarray}
&& A^{*}_{u}(s, d(n))+A^{*}_{l}(s, d(n)) \le A^{*}_{u}(s, d(n)) + l^{M_l} \nonumber\\
&\le& A_{u} (s, d(n)-\frac{l(n)}{c})+ l^{M_l} \nonumber\\
&\le& \rho_u(d(n)-s-\frac{l(n)}{c})) + \sigma_u+ l^{M_l}
\end{eqnarray}
which, when applied to (\ref{eq-3aa}), results in
\begin{eqnarray}
d(n) &\le &  s + \frac{\sum_{k=n_0}^{n} l(k)}{c} + \frac{\rho_u (d(n)-\frac{l^{m_i}}{c} -s) + \sigma_u+  l^{M_l}}{c}  \nonumber 
\end{eqnarray}
Further with simple manipulation, we obtain
\begin{equation}\label{eq-6a}
d(n) \le s + \frac{\sum_{k=n_0}^{n} l(k)}{c-\rho_u} + \frac{\sigma_u + l^{M_l} - \frac{\rho^u}{c}l^{m_i}}{c-\rho_u} 
\end{equation}
and with $s \le a(n_0)$, we have 
\begin{eqnarray}
d(n) &\le& a(n_0)+ \frac{\sum_{k=n_0}^{n} l(k)}{c-\rho_u} + \frac{\sigma_u + l^{M_l} - \frac{\rho^u}{c}l^{m_i}}{c-\rho_u} \nonumber \\
&\le& \max_{0\le m \le n}\{ a(m)+ \frac{L(m,n)}{c-\rho_u} +E \} + \frac{ l(n)}{c-\rho_u}  \label{eq-6a}
\end{eqnarray}
which completes the proof of part (i). With part (i), part (ii) immediately follows from Corollary \ref{cor-2}.  
\end{proof}
 
\vspace{6pt}
\noindent{\bf C. Proof of Theorem \ref{th-cbs4}} 
\begin{proof}
To illustrate credit-frozen, Figure  \ref{fig-cbs4} is presented. 

\begin{figure}[h!]
\centering
  \includegraphics[width=0.95\linewidth]{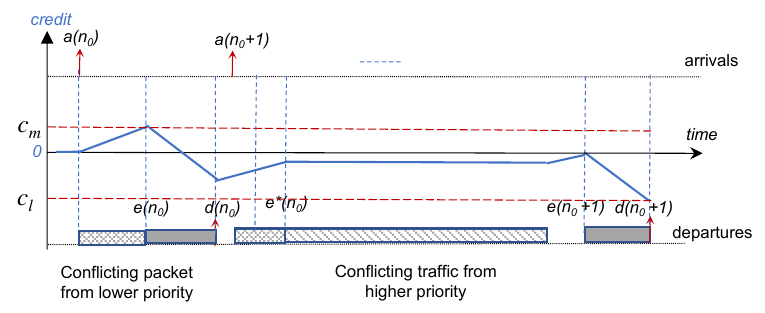} 
  \caption{$Credit$ function frozen during transmission of higher priority traffic} 
  \label{fig-cbs4}
\end{figure}

For part (i), we shall follow the approach used in proving Theorem \ref{th-cbs2}. 
For any packet $n (\ge 1)$ from the considered CBS queue, let $e(n)$ denote the time when the packet enters transmission. In addition, let $n_0 ( \le n)$ be $n_0 = \max\left\{m (\le n): credit (a(m))=0 \textrm{ and } a(m)>d(m-1)\right\}$. 

Let $\Delta t\equiv d(n) - a(n_0) $, and let $\Delta t^{\uparrow}$ , $\Delta t^{\downarrow}$ and $\Delta t^{\rightarrow}$ respectively denote the accumulated length of all periods in $[a(n_0), d(n)]$ when $credit$ is increasing, decreasing or frozen respectively. Since the holding periods in $\Delta t$ are only due to higher priority traffic, we have 
\begin{eqnarray} 
\Delta t &=& \Delta t^{\uparrow} +  \Delta t^{\downarrow} + \Delta t^{\rightarrow}  \label{eq6-1}
\end{eqnarray}

Since the decreasing periods are the transmission periods of packets $n_0$ to $n$, we have 
\begin{eqnarray} 
c\cdot \Delta t^{\downarrow} &=&  \sum_{m=n_0}^{n}l(m) = L(n_0, n+1)  \label{eq6-2}
\end{eqnarray}
In addition, with $credit (a(n_0))=0$ by the definition of $n_0$ and the changing rate is zero during the credit on-hold periods, we have
\begin{eqnarray} 
&& credit(d(n)) \nonumber \\
 &=& credit(a(n_0))+ I\cdot \Delta t^{\uparrow} + S \cdot \Delta t^{\downarrow} + 0 \cdot \Delta t^{\rightarrow}  \nonumber \\
&=& I\cdot \Delta t^{\uparrow} + S \cdot \Delta t^{\downarrow} \nonumber \\
&=& I\cdot \Delta t^{\uparrow} + \frac{S \cdot L(n_0, n+1)}{c} \nonumber 
\end{eqnarray}

Since the value of $credit$ is frozen during the transmissions of high priority packets, the same maximum and minimum values under the setting of having CBS at the highest priority level will apply, which is the setup for Theorem \ref{th-cbs2}, where, we have proved in (\ref{eq-credit-n})
\begin{eqnarray}
credit(d(n)) &\le &  I \frac{l^{M_l}}{c} + S \frac{l(n)}{c} \equiv credit^{M}\label{eq-credit-na}
\end{eqnarray}
We now have  
$$I\cdot \Delta t^{\uparrow} + \frac{S \cdot L(n_0, n+1)}{c} \le \frac{I \cdot l^{M_l}}{c}+ S \frac{l(n)}{c} $$
and hence 
\begin{eqnarray} 
c\cdot \Delta t^{\uparrow}  &\le& \frac{- S \cdot L(n_0, n+1)}{I} + l^{M_l} + S \frac{l(n)}{I}  \nonumber \\
&=& \frac{- S \cdot L(n_0, n)}{I} + l^{M_l}  \label{eq6-3}
\end{eqnarray}

Moreover, the high priority traffic that has frozen $credit$ in periods within $[a(n_0), e(n)]$ must have all arrived in $(e(n_0), e(n) - \frac{l^m}{c}$). This is because at $e(n_0)$, if there were a high priority packet in the system, the CBS packet $n_0$ would not have been able to enter transmission. In addition, the last high priority packet must have finished transmission by $e(n)$ such that $n$ can start at $e(n)$. Since the transmission of such a high priority packet takes at least $\frac{l^{m_u}}{c}$, it must have arrived at least before $e(n) - \frac{l^{m_u}}{c}$. 
Since the high priority traffic is token bucket constrained, there holds
\begin{eqnarray} 
c \cdot \Delta t^{\rightarrow} &\le& A(e(n_0), e(n)- \frac{l^{m_u}}{c})  \nonumber \\
&\le& \rho_{u}(e(n)-e(n_0)- \frac{l^{m_u}}{c}) +\sigma_{u} \nonumber \\
&\le& \rho_{u}(e(n)-a(n_0)- \frac{l^{m_u}}{c}) +\sigma_{u}\nonumber \\
&=& \rho_{u}(\Delta t- \frac{l^{m_u}}{c} - \frac{l(n)}{c}) +\sigma_{u} \label{eq6-4}
\end{eqnarray}
where we have applied $e(n) = d(n) - \frac{l(n)}{c} $ and $\Delta t = d(n)-a(n_0)$. 

Applying (\ref{eq6-2}),  (\ref{eq6-3}) and  (\ref{eq6-4}) to  (\ref{eq6-1}), we can get
\begin{eqnarray} 
c \Delta t &\le& \frac{- S \cdot L(n_0, n)}{I} + l^{M_l}  + L(n_0, n+1) \nonumber \\
&& +  \rho_{u}(\Delta t- \frac{l^{m_u}}{c} - \frac{l(n)}{c}) +\sigma_{u} \nonumber \\
&\le&  \frac{c \cdot L(n_0, n)}{I} + l^{M_l}  + l(n) \nonumber \\
&& +  \rho_{u}(\Delta t- \frac{l^{m_u}}{c} - \frac{l(n)}{c}) +\sigma_{u} \nonumber \\
\end{eqnarray}
from which, we further obtain
\begin{eqnarray} 
\Delta t &\le& \frac{c \cdot L(n_0, n)}{I(c- \rho_{u})} + \frac{\sigma_{u} + l^{M_l}}{c- \rho_{u}} + \frac{l(n)}{c} - \frac{l^{m_u} \rho_{u}}{c(c- \rho_{u})}  \nonumber \\
 &\le& \frac{c \cdot L(n_0, n)}{I(c- \rho_{u})} + \frac{\sigma_{u} + l^{M_l}}{c- \rho_{u}} + \frac{l(n)}{c} 
 \end{eqnarray}
and since $\Delta t = d(n) - a(n_0)$, we obtain
\begin{eqnarray} 
&&d(n)  \nonumber\\
&\le& a(n_0) + \frac{c \cdot L(n_0, n)}{I(c- \rho_{u})} + \frac{\sigma^{u} + l^{M_l}}{c- \rho_{u}} + \frac{l(n)}{c} \nonumber\\
&\le& \max_{0\le m \le n}\{ a(m) + \frac{L(m, n)}{R} + \frac{\sigma_{u} + l^{M_l}}{c- \rho_{u}} \} + \frac{l(n)}{c} \label{eq7-1} \\
&\le& \max_{0\le m \le n}\{ a(m) + \frac{L(m, n)}{R} + \frac{\sigma_{u} + l^{M_l}}{c- \rho_{u}}  \nonumber\\
&& - (\frac{1}{R}-\frac{1}{c})l^{m_i}\} + \frac{l(n)}{R} \label{eq7-2} 
\end{eqnarray}
with $R= \frac{c- \rho_{u} }{c}I$. From (\ref{eq7-1}), the $g_1$-server part of (i) is proved.  The $g_2$-server part of (i) follows from  (\ref{eq7-2}). 

With the $g_2$-server part of (i), part (ii) follows immediately from Corollary \ref{cor-2}.
\end{proof}

\end{document}